\documentclass[conference, anonymous]{IEEEtran}
\IEEEoverridecommandlockouts
\usepackage{cite}
\usepackage{amsmath,amssymb,amsfonts}
\usepackage{algorithmic}
\usepackage{graphicx}
\usepackage{textcomp}
\usepackage{multirow}
\usepackage{booktabs}
\usepackage{threeparttable}
\usepackage{enumitem}
\usepackage[framemethod=tikz]{mdframed}
\usepackage{tikz}
\usepackage{tipa}
\usepackage{xtab}
\usepackage{mfirstuc}
\usepackage{todonotes}
\usepackage{soul}
\usepackage{colortbl}
\usepackage{xspace}
\usepackage{xcolor}
\usepackage{url}

\usepackage{caption}
\usepackage{subcaption}
\usepackage{colortbl}
\usepackage[many]{tcolorbox}
\newtcolorbox{mybox}{
  breakable,
  colback=white,
  colbacktitle=white,
  coltitle=black,
  bottomrule=0pt,
  toprule=0pt,
  leftrule=3pt,
  rightrule=3pt,
  titlerule=0pt,
  arc=0pt,
}

\usepackage{booktabs}\usepackage{dirtytalk}  
\MakeRobust{\say} 

\newboolean{showcomments}
\setboolean{showcomments}{true} 
\ifthenelse{\boolean{showcomments}}
    {

        \newcommand{\zane}[1]{\todo[inline,linecolor=blue,backgroundcolor=blue!25,bordercolor=blue]{\textcolor{black}{\textbf{Zane says:}} #1}}
        \newcommand{\nowshin}[1]{\todo[inline,linecolor=red,backgroundcolor=red!25,bordercolor=red]{\textcolor{black}{\textbf{Nowshin says:}} #1}}
        \newcommand{\kezia}[1]{\todo[inline,linecolor=cyan,backgroundcolor=yellow!25,bordercolor=purple]{\textcolor{black}{\textbf{Kezia says:}} #1}}
        \newcommand{\manish}[1]{\todo[inline,linecolor=green,backgroundcolor=green!25,bordercolor=green]{\textcolor{black}{\textbf{Manish says:}} #1}}
        \newcommand{\amanda}[1]{\todo[inline,linecolor=yellow,backgroundcolor=orange!25,bordercolor=black]{\textcolor{black}{\textbf{Amanda says:}} #1}}
        \newcommand{\neil}[1]{\todo[inline,linecolor=yellow,backgroundcolor=yellow,bordercolor=black]{\textcolor{black}{\textbf{Neil says:}} #1}}
        \newcommand{\dana}[1]{\todo[inline,linecolor=orange,backgroundcolor=pink,bordercolor=black]{\textcolor{black}{\textbf{Dana says:}} #1}}
    }
    {
        \newcommand{\zane}[1]{}
        \newcommand{\nowshin}[1]{}
        \newcommand{\kezia}[1]{}
        \newcommand{\manish}[1]{}
        \newcommand{\amanda}[1]{}
        \newcommand{\dana}[1]{}{}
        \newcommand{\neil}[1]{}{}
    }

\newcommand{\qs}[2]{\emph{``#1'' (#2)}}
\newcommand{\qsd}[3]{\emph{``#1'' (#2 on #3)}}

\begin{document}
\thispagestyle{plain}
\pagestyle{plain}
\title{A Data-Driven Approach for Finding Requirements
Relevant Feedback from TikTok and YouTube \\}


\author{\IEEEauthorblockN{
Manish Sihag\textsuperscript{\textsection},
Ze Shi Li\textsuperscript{\textsection}, 
Amanda Dash,
Nowshin Nawar Arony,
Kezia Devathasan,
Neil Ernst, \\
Alexandra Branzan Albu,  
Daniela Damian}
\IEEEauthorblockA{
\emph{Department of Computer Science}\\
\emph{University of Victoria, Victoria, Canada}\\
\emph{\{manishsihag, lize, adash42, nowshinarony, keziadevathasan, nernst, aalbu, danielad\}@uvic.ca}}}

\maketitle
\begingroup\renewcommand\thefootnote{\textsection}
\footnotetext{The first two authors contributed equally to this work}
\endgroup

\begin{abstract}

The increasing importance of videos as a medium for engagement, communication, and content creation 
makes them critical for organizations to consider for user feedback.
However, sifting through vast amounts of video content on social media platforms to extract requirements-relevant feedback is challenging. 
This study delves into the use of TikTok and YouTube, two widely used social media platforms that focus on video content, in identifying relevant user feedback that may be further refined into requirements using subsequent requirement generation steps.
We demonstrate an approach of using videos as a source of user feedback by analyzing  audio and visual text, and metadata (i.e., description/title) from 6276 videos of 20 popular products across various industries. 
We employed state-of-the-art deep learning transformer-based models, and classified 3097 videos consisting of requirements relevant information.
We then clustered relevant videos and found multiple requirements relevant feedback themes for each of the 20 products.  
This feedback can later be refined into requirements artifacts. 
We found that product ratings (feature, design, performance), bug reports, and usage tutorial are persistent themes from the videos. 
Video-based social media such as TikTok and YouTube can provide valuable user insights, making them a powerful and novel resource for companies to improve customer-centric development.

\end{abstract}


\section{Introduction}
Online videos are becoming more important for organizations to consider for user feedback as videos provide an immersive experience for viewers.
Videos are a very popular medium for social media and communication \cite{xu2019research}. 
For example, TikTok is one of the world's most popular video-based social media platforms \cite{xu2019research, clarissa2022rising} and
YouTube has also grown to an astronomical magnitude \cite{hoffman2012toward}.

Previous research on requirements and videos has been limited to investigating the comments section of videos, while users engage in discussion~\cite{das2019youtube, madden2013classification, karras2021potential}. 
However, videos are rich sources of data \cite{khanresearching} with  both audio and visual components, and metadata (e.g., description, title, date created). 
In this study we look at all three sources: the audio track of the video (converted to a transcript), any text that appears in the video, such as captions and subtitles, and the metadata.

Paying attention to the direction of CrowdRE research is critical for companies to improve requirements elicitation \cite{groen2015towards, groen2017crowd}. 
The ability to vastly increase the amount of feedback considered \cite{groen2016requirements} is extremely valuable. 
The process we propose converts video content to requirements-relevant feedback that can significantly impact companies' requirements and development activities.

We present a data-driven exploratory study on leveraging user-generated videos from TikTok and YouTube to identify requirements-related user feedback for 20 distinct products. 
This information can serve as a foundation for requirement elicitation, facilitating a more comprehensive understanding of consumer preferences and needs. 
We analyzed videos about products from a variety of industries, including software, consumer electronics, and automotive.
Our approach involves extracting textual data from audio and visual content from the videos and processing using natural language processing (NLP) and machine learning (ML) techniques to uncover important user feedback that may not be captured through traditional elicitation methods.

Our study contributes to the growing body of research on using social media as a data source for product development and user feedback analysis. 
It also provides insights into the strengths of using videos as a data source and the opportunities of applying NLP and ML techniques to analyze video data.
Our study was guided by three central research questions:
\begin{enumerate}[label=\textbf{RQ\arabic*},leftmargin=*]
    \item How can video-based social media be used to identify requirements relevant user feedback?
        
    \item What are the main users feedback themes that we can identify?
    \item How do the different social media platforms and their video content affect user feedback?  
\end{enumerate}

From this exploration of videos contents from TikTok and YouTube, a number of findings have emerged.
Our study presents the following contributions:

\begin{itemize}
    \item An approach for identifying requirement relevant user feedback from video based social media. 
    \item The most effective machine learning models (GPT-2 and RoBERTa) to classify the audio and visual content into requirements relevant user feedback.
    \item A list of requirements relevant user feedback themes for software, phone, computer, and automotive industries which can be further refined into requirements. 
\end{itemize}

\section{Background and Related Work} 
\subsection{CrowdRE in Requirements Elicitation}
With the rapid evolution of technology and social media, traditional elicitation techniques are insufficient to identify, gather and formulate requirements from the large distributed online community~\cite{groen2015towards}. 
To address this gap, Groen \emph{et al.} \cite{groen2015towards} proposed CrowdRE
\textit{``a semi-automated requirement engineering (RE) approach for obtaining and analyzing any kind of `user feedback' from a `crowd', with the goal of deriving validated user requirements.''}
User feedback from the `crowd' is then transformed into requirements either through manual content analysis \cite{gebauer2008user} or through automated approaches \cite{jiang2014user}. 
Groen \emph{et al.} argue that CrowdRE can address the limitations of traditional RE methods, such as the limited scope and representation of user feedback \cite{groen2017crowd}.
By harnessing the collective intelligence of a crowd, organizations can utilize CrowdRE to identify and prioritize user needs and improve user engagement for their product\cite{wang2019systematic}. 

The services offered by CrowdRE aim to provide motivational tools (e.g. gamification techniques, forums, visuals) that can inspire stakeholders to actively participate in a crowd \cite{groen2016requirements}. 
A number of studies have focused on leveraging this crowd engagement on various platforms like app reviews and forums \cite{tizard2019can, maalej2015bug, chen2014ar, pagano2013user, di2016would}, demonstrating that valuable insights can be gained by analyzing the conversations generated by users, such as their comments, feedback, and suggestions. 
Moreover, social media platforms have also been studied and utilized to analyze various aspects of requirement engineering \cite{kanchev2015social, kengphanphanit2020automatic, williams2017mining}.
Li \emph{et al.} \cite{li2022narratives}, found privacy related user feedback in a study on product related subreddits in Reddit. 
Kengphanphanit  \emph{et al.} \cite{kengphanphanit2020automatic}, classified user feedback into requirements and non-requirements by scraping Twitter and Facebook, and utilized feature extraction on the user feedback based on polarity, subjective, and number of requirements word factors. 
Afterwards, they developed a model using the three factors and Naive Bayes method to generate requirements from the user feedback.  
Nevertheless, few studies have focused on exploring video based social media platforms such as YouTube and TikTok for identifying requirements relevant user discussions that have potential to be refined as requirements later on. 

\subsection{Video Platforms}
YouTube with over 2.5 billion active users and TikTok with over 1 billion users, are the most popular video based social media platforms \cite{noauthor_biggest_nodate}. 
Video platforms like YouTube and TikTok have emerged as a valuable source for user engagement and obtaining requirements relevant user discussions \cite{das2019youtube, madden2013classification, vistisen2017return}. 
In a paper by Madden \emph{et al.} \cite{madden2013classification}, the authors conducted an analysis on the content from 66,637 YouTube comments, and found 10 broad categories of user discussion, suggesting that classifying YouTube comments revealed opinions and attitudes of viewers towards the video content.
In their study Das \emph{et al.} \cite{das2019youtube} analyzed the comments generated on YouTube videos using natural language processing techniques and categorized comments on YouTube videos related to autonomous vehicles, further suggesting that YouTube can be a useful source of information for understanding consumer opinions and concerns.
Karras \emph{et al.} \cite{karras2021potential} utilized machine learning algorithms to analyze 4505 comments from a YouTube video as source of feedback and classified them into product relevant comments. 
The further manually analyzed the content of the relevant comments and found discussion on feature request, problem report, efficiency, and safety from the product relevant comments. 

Schneider \emph{et al.} \cite{schneider2019video}, in their study describe that different types of videos (linear videos, vision videos, and interactive videos) can demonstrate concrete situations regarding a product and can be beneficial in engaging users to solicit feedback. 
Vision videos depict a vision of a future product or system, and they can help stakeholders to better understand and communicate their needs \cite{karras2021supporting}.
Hence, studies have been conducted on leveraging vision videos to elicit user feedback as users frequently engage in discussions and provide feedback on these videos \cite{schneider2019refining, busch2022vision}. 
In another study, Karras \emph{et al.} \cite{karras2021potential}, argue that although the existing literature \cite{busch2020vision, schneider2019refining} on vision videos discusses the benefit of using them in soliciting feedback, its potential for CrowdRE has not yet been fully explored. 
It remains unclear whether videos created by content creators themselves can provide valuable insights for companies.

Thus, in our study, we explore the potentials of video contents from YouTube and TikTok for identifying user feedback.
Our work focuses on finding the pertinent themes from the user feedback.
Since the information from the `crowd' may also generate irrelevant data, extracting relevant information is a critical step before creating requirements.
Once relevant themes have been identified, companies may use manual \cite{gebauer2008user} or automated approaches \cite{kengphanphanit2020automatic} to generate requirements from the themes.  

\section{Methodology}
\begin{figure}[h!]
  \centering
  \includegraphics[width=1\linewidth]{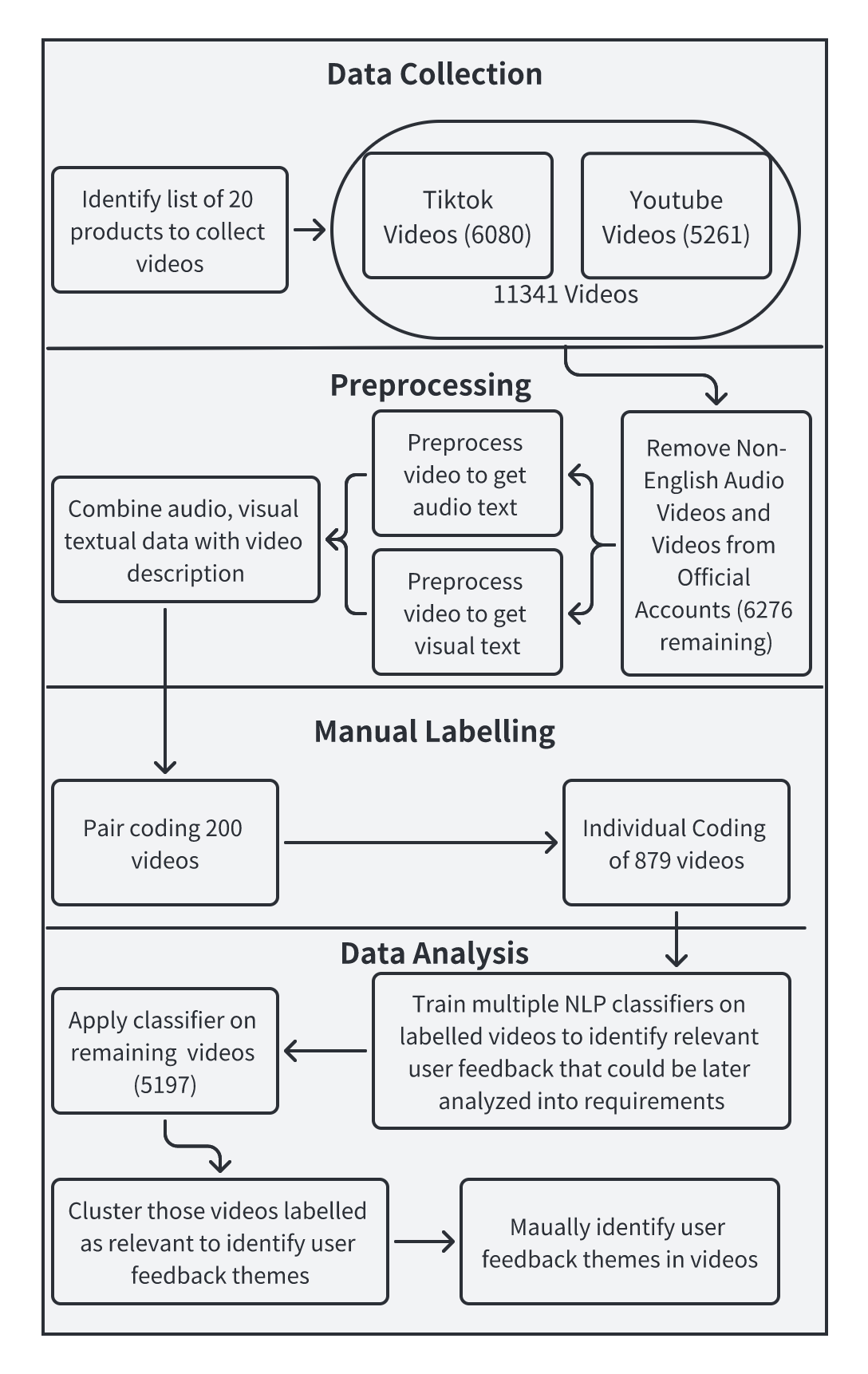}
  \caption{Research Process}

  \label{research-process}

\end{figure}

We conducted an exploratory study to investigate the feasibility of using video-based social media platforms (i.e., TikTok and YouTube) for identifying requirements relevant user feedback themes. 
Our methodology is summarized by Figure \ref{research-process}.

\subsection{Data collection}
We conducted extensive market research and analysis to identify the top-performing products in each industry to build a representative dataset of twenty different products from 4 common consumer categories: software, mobile phone, computer, and automotive.
Table \ref{20-products} shows the dataset characteristics.
We chose the most widely used software across different domains, including browsers such as Chrome and Firefox, tutoring applications like Duolingo, networking platforms like Discord, and productivity software like Notion.
For each of the other categories, we strived to pick products that were among the best selling flagship products in North America in 2022, with on focus videos that would be in English. 
We chose not to select products that may otherwise sell more units overall worldwide, such as Vivo versus Oneplus, but have a smaller English audience as they would impact the videos that we could collect. 
For example, for the products from the automotive industry, we chose 5 brands and their best selling model in North America in 2022.
We applied the same approach to mobile phones and computers by selecting the most popular flagship product released in 2022.

To collect the data, we used the public facing APIs from TikTok and YouTube and scraped the videos by searching for each product using its name.
The search term for each product is provided in the replication package \cite{manish_sihag_2023_8088427}.
We downloaded all the available videos from each search term and this process took about a day for each product. 
In total, we collected 11,341 videos, with 6,080 from TikTok and 5,261 from YouTube.

\begin{table}[h]

\caption{Products used for the analysis}
\centering
\small
\begin{tabular}{p{1.3cm}p{3.3cm}p{1.3cm}p{1.3cm}}
\toprule

\textbf{Category} &  \textbf{Products} & \textbf{TikTok Videos}  &  \textbf{YouTube Videos} \\ \midrule
\multirow{5}{*}{Software}  & Notion & 280 &  232 \\
 & Duolingo & 224 &  217 \\
& Discord & 94  &  103 \\
 & Chrome & 82 & 105 \\
 & Firefox  & 50 & 189 \\
 \midrule
\multirow{5}{*}{Phone}  & Google Pixel 7 & 223 & 183 \\
    & Apple Iphone 14 & 178 & 142 \\
    & Samsung Galaxy S22 & 162 & 214 \\
    & Motorola Edge 30 & 76 & 92 \\ 
    & Oneplus 10  & 59 & 119 \\ \midrule
    \multirow{5}{*}{Computer}  & Microsoft Surface Pro 9  &  201 & 187 \\
& Apple Macbook Air M2 & 193 & 161 \\
 & Asus Zenbook 14 & 119 & 132 \\
 & HP spectre x360 14 & 130 & 95 \\
 & Dell XPS 15 & 30 & 49 \\ \midrule
\multirow{5}{*}{Automotive}  & Tesla Model 3 & 210 & 193 \\
  & BMW X5 & 190 & 197 \\
 & Ford F150 & 187 & 102 \\
 & Toyota Rav4 & 177 & 305 \\
 & Mercedes Benz GLC & 154 & 239\\ 

\bottomrule
\end{tabular}

\label{20-products}
\end{table}

\subsection{Preprocessing}

The videos that we collected represented all the available videos according to our search term for each product.
To ensure that our dataset is focused on user-generated content and not official promotional material, we implemented a two-level data filtration process. 
First, we filtered out any videos uploaded by official product accounts, as they are more likely to discuss product features in a promotional manner.
The remaining videos in our dataset represented those that matched our search terms, but not from official product accounts such as Apple.
Second, we filtered the videos to only include those in the English language. 
We used Spacy FastLang \cite{spacy_fastlang} to detect the language of the video description text, and OpenAI Whisper \cite{radford_robust_nodate} to detect the language from the audio text, as described in detail in Section \ref{text_from_audio}.
We were left with 6276 videos after filtration.

\subsection{Analysis of Videos}
Videos contain audio tracks, metadata in the form of descriptions, and finally text that appears in the video itself (such as a caption or subtitle).
We used both the audio and visual text data extracted from the videos, as well as the descriptions provided by the content creators. 
To make sense of videos we worked with both the \emph{visual} and \emph{audio} elements of a video.
First, we converted the audio of the video into text; secondly, we sampled visual frames and performed computer vision to collect any displayed text in a video (i.e., video subtitles).
Out of the total of 6,276 videos in our dataset, 403 videos were found to have no audio content. Additionally, 101 videos did not contain any visual text.
For each video we also paired this textual content with a video's metadata including video description and title where applicable. We then classified the TikTok and YouTube videos using various state-of-the-art deep learning models as user feedback that could be later refined into requirements (referred to as ``relevant") or user feedback that was not useful for later refinement into requirements (``irrelevant"). 
We describe these techniques below.

\subsubsection{Extracting Text from the Audio} \label{text_from_audio}
OpenAI's speech recognition model ``Whisper"  \cite{radford_robust_nodate} was used to extract audio text from videos.
The ``Large" Whisper model used in this study is one of the most accurate models and is designed for high-quality transcription tasks.

The extracted audio from videos was run on the Whisper model to generate transcriptions of the audio content.
On average each TikTok videos was less a minute in duration, where as the YouTube videos were roughly eight minutes in length. 
Hence, some YouTube videos were a little bit longer in length. To reduce computational time, for the longer YouTube videos, we transcribe only the first thirty minutes of each audio and trim the rest. 
We assume that the premise of the video is conveyed in first 30 minutes, to minimize the processing time for longer videos. 
Whisper can detect the audio language being spoken during the transcription process and we utilized this to filter out videos that were not in English, as it is imperative to ensure the accuracy and relevance of the extracted text for our analysis.

\subsubsection{Extracting Visual Text from the Video}

In addition to audio, we also extracted visual text that may appear in a video as some content creators display subtitles or other important visual text.  Since videos consist of many repetitive or similar frames, \textit{motion-based video summarization} is used to select a small subset of all frames. For frames that are detected as having ``text", we use an optical character recognition (OCR) system to extract the text; common spelling errors are corrected.  The following section describes the video extraction pipeline.


For candidate frame selection, we use a modified version of the algorithm proposed by Dash and Albu\cite{dash2017domain}.  This algorithm was chosen because it is a heuristic and not a ML-based video summarization system, therefore it is independent of the video domain. Their approach integrates motion and saliency analysis with temporal slicing to extract features and unique candidate frames from the video.

We do not use their candidate frame summarization; instead we leverage the information from the saliency energy map (instead of the slower background subtraction model) to find the probability divergence for the temporal slices.  We take the Kullback-Leibler divergence, $D_{KL}(.)$, of each temporal slice, $k \in \{\text{vertical}, \text{horizonal}, \text{diagonal}\}$ at time $t-1$ and $t$, where $t$ is defined as the current frame.  We thus obtain a vector $s_{t} \in \mathbb{R}^{3}$ (Eqn.~\ref{eqn:kl}),  
\begin{equation}\label{eqn:kl}
s^{(k)}_{t} = D_{KL} ( p(k)_{t} || p(k)_{t-1} ) \\
\end{equation}

where $p(k)$ is the temporal slice $k$, normalized as a probability vector. 
The vector is then thresholded by values greater than $T_h \in \mathbb{R}^3$ to select the candidate frame (Eqn.~\ref{eqn:candidate}); for this paper $T_h = [1e-4, 1e-4, 1e-4]$ is used.  
\begin{equation}\label{eqn:candidate}
     \text{candidate frame} =
     \begin{cases}
         f_t & \forall k \in \{(s^{(k)}_{t} - s^{(k)}_{t-1}) > T_h\}, \\
         \emptyset & \text{otherwise}
     \end{cases}
 \end{equation}

Intuitively, when the movement distribution changes significantly, a new candidate frame is selected.  The candidate frames are analyzed for text using CentripetalText~\cite{sheng2021centripetaltext}. If no sufficient size text is detected, the candidate frame is discarded.  
In the next step, we consider two scenarios: (1) an audio track exists, and (2) no audio track exists.

When the primary content in a video is visual (i.e. no audio), we utilized Google's commercial state-of-the-art OCR system called ``Google Cloud Vision"\footnote{https://cloud.google.com/vision} to capture all text in the candidate frame.  When audio is available, we assume the video content is primarily communicated by audio.  Therefore, we supplement the audio by using HuggingFace's open-source OCR Tr-OCR~\cite{li2021trocr} system with the ``trocr-large-printed` pretrained weights to extract larger OCR text discovered by CentripetalText. Both OCR methods are not completely accurate, so we attempt to fix common spelling mistakes by processing the raw extract text with Peter Norvig's algorithm.\footnote{https://norvig.com/spell-correct.html} The choice to use multiple OCR algorithms was due to budget constraints.  

\subsection{Manual Labeling}
To evaluate the accuracy and effectiveness of our classification models, we employed a manual labeling process to create a ground truth dataset for training. 
The dataset was randomly selected from our entire data pool, and we labeled total 1079 videos. 
Our labeling process consisted of labeling videos as either ``relevant"
or ``irrelevant".

Our criteria to labelling a video as ``relevant" include aspects such as problem reports, reviews of a product feature, comparison of features with competitors, feature requests, etc.
In other words, any time a video included content that could be used by a company to make informative decisions regarding changes (positive or negative) to their product, we labelled it as ``relevant".
For example, \qs{To find out what the safest browser to use in 2022 is based on empirical testing techniques So we re going to go through 200 of the latest malware links... Firefox only blocked 145... Chrome not quite as good as Edge it blocked 198 links out of 200...}{Firefox} was labeled as relevant. 
In contrast, a video that do not describe a product in any meaningful way or in a superficial manner (i.e., ``The new M2 MacBook Air is finally for sale. I’m not gonna buy one") was labelled as irrelevant.
Exemplified by these quotes, the pair coders would label a video as ``relevant" if the main point of the video or substantial part of the video content (i.e., several sentences with details) discusses the product in a way that a company could take actionable steps.
In the case of short videos, where the total amount of content is just a few sentences, the threshold for labeling as ``relevant" could be a single sentence, but the sentence would need to provide adequate details.


To prevent bias towards any specific product, we made sure to label videos from each product that we analyzed.
Two of our authors with extensive experience in requirement analysis, pair coded a set of 200 videos for the manual labeling process. 
The pair coding process resulted an average Cohen's Kappa score of 87\%, indicating high levels of agreement between the coders.
This high inter rater reliability also indicated that the separation between ``relevant" and ``irrelevant" was quite clear.
After the successful completion of the pair coding process, one author individually labeled the remaining 897 videos. Of the 1079 videos that were manually labeled, 601 were labeled as relevant and 478 were labeled as irrelevant.

\subsection{Data Analysis of User Feedback}
\subsubsection{Classification}
We employed five state-of-the-art deep learning transformer-based models, namely GPT-2 (Generative Pre-trained Transformer 2)\cite{radford_robust_nodate}, BERT (Bidirectional Encoder Representations from Transformers)\cite{devlin_bert_2019}, RoBERTa (Robustly Optimized BERT Approach)\cite{liu_roberta_2019}, XLM-RoBERTa (Cross-lingual Language Model - Robustly Optimized BERT Approach)\cite{conneau_unsupervised_2020}, and ALBERT (A Lite BERT) \cite{lan_albert_2020}, to classify videos as either relevant or irrelevant. 
Fine-tuning these models allowed us to identify the most effective approach for video classification.

We evaluated the performance of these models using different combinations of data, including visual text, audio text, and both audio and visual text. Furthermore, we included title and description data for all combinations. 
By comparing the performance of these models, we aimed to identify the optimal approach for accurately classifying textual data from these popular video sharing platforms. 
For each of the five deep learning models (GPT-2, BERT, RoBERTa, XLM-RoBERTa, and ALBERT), we followed a similar training process. We used the pre-trained models and fine-tuned them on our dataset of labeled video text data, which included both the audio and visual text, as well as the video metadata (i.e., title and description).
After training, we evaluated the performance of each model on a balanced test set of video text data.
We measured the performance using accuracy and area under curve (AUC) metrics. 
We repeated this process for each combination of data (visual text, audio text, and both audio and visual text) and for each platform (YouTube and TikTok) to compare the performance of the models on each type of data and platform.
 
\subsubsection{Clustering}
We clustered the data to learn user feedback themes. 
We used BERTopic \cite{grootendorst_bertopic_2022} to infer documents distribution over topics and then use BERTopic topic descriptions for clustering.
BERTopic allows us to chose the cluster model. We selected K-means as our cluster model and ran the clustering process for 2 to 6 clusters. 
To determine the best cluster, we used the Silhouette Coefficient \cite{rousseeuw_silhouettes_1987}, a metric that measures how similar an object is to its own cluster compared to other clusters. 
After forming clusters, we conducted a manual analysis to thoroughly review the formed clusters. Based on that, we created theme names and assigned them to each cluster to represent their respective content.

\begin{table}[th!]

    \centering
    \caption{Results of Different Deep Learning Models on Classifying between Relevant vs Irrelevant. AUC is area under curve. }
    \begin{tabular}{p{2.7cm}p{1.9cm}cc} \toprule
      Dataset & Model & Accuracy & AUC \\\midrule
    \multirow{5}{*}{\shortstack[c]{YouTube with only \\ visual text}} & GPT-2 & 0.71  & 0.71  \\ 
     & BERT & 0.76  & 0.76  \\ 
     & RoBERTa & 0.74  & 0.74 \\ 
     & XLM-RoBERTa & 0.67  & 0.67  \\ 
     & ALBERT & \textbf{0.79}  & \textbf{0.79}  \\   \midrule
    \multirow{5}{*}{\shortstack[c]{YouTube with only \\ audio text}} & GPT-2 & \textbf{0.94}  &  \textbf{0.94} \\ 
     & BERT & 0.86 & 0.86   \\ 
     & RoBERTa & 0.86 & 0.86 \\ 
     & XLM-RoBERTa & 0.83 &  0.83\\ 
     & ALBERT & 0.79  & 0.79 \\ 
        \midrule
     \multirow{5}{*}{\shortstack[l]{YouTube with both \\ visual  and audio text}} & GPT-2 & \textbf{0.91}  &\textbf{ 0.91}  \\  
     & BERT & 0.85  & 0.85   \\   
     & RoBERTa & 0.80  & 0.80  \\   
     & XLM-RoBERTa & 0.80  & 0.80  \\   
     & ALBERT& 0.79  & 0.79  \\    \midrule
    \multirow{5}{*}{\shortstack[c]{TikTok with  only \\ visual text}} & GPT-2 & \textbf{0.71}  & \textbf{0.71}   \\  
     & BERT & 0.70  & 0.70  \\  
     & RoBERTa & 0.50  & 0.50  \\  
     & XLM-RoBERTa & 0.50  & 0.50   \\   
     & ALBERT & 0.70  & 0.70   \\    \midrule
    \multirow{5}{*}{\shortstack[c]{TikTok  with only \\  audio text}}
     & GPT-2 & 0.92 & 0.92\\ 
     & BERT 0.92 & 0.92 & 0.92  \\ 
     & RoBERTa & \textbf{0.93} & \textbf{0.93} \\ 
     & XLM-RoBERTa & 0.90 &  0.90  \\ 
     & ALBERT & 0.90 &  0.90  \\ \midrule
    \multirow{5}{*}{\shortstack[c]{TikTok with both \\ visual and audio text}}  & GPT-2 & 0.93 &   0.93 \\ 
     & BERT & 0.95 &  0.95 \\ 
     & RoBERTa & \textbf{0.97} &   \textbf{0.97} \\ 
     & XLM-RoBERTa & 0.90  & 0.90 \\ 
     & ALBERT & 0.93  & 0.93 \\
    
    \end{tabular}

    \label{tab:train_test_models}
\end{table}




  
 
   
    
    

\section{Findings}
\subsection{RQ1: How can video-based social media be used to identify requirements relevant user feedback?}


Recall that for each video we  1) converted the audio track into text and 2) performed computer vision to collect any displayed text in a video (i.e., captions or video subtitles).
We then classified the TikTok and YouTube videos using various state-of-the-art deep learning models as either ``relevant"  or ``irrelevant"; 
The results of the classification using these techniques are summarized in Table~\ref{tab:train_test_models}.
We observe that the datasets that leverage \emph{audio} text paired with video metadata always performs extremely well.
In particular, \emph{audio} text paired with video metadata consistently performed better than \emph{visual} text paired with video metadata.
For YouTube videos and TikTok videos that utilize \emph{audio} text paired with video metadata, accuracies of 94\% and 93\% were achieved.

We contrast these results with those datasets that leveraged visual text paired with video metadata.
Table \ref{tab:train_test_models} shows that solely relying on text extracted from the video frames is not sufficient to identify requirements relevant user feedback.
The most accurate model for classifying the dataset for YouTube's visual text was only able to achieve an accuracy equal to the worst performing model for the YouTube audio text dataset.
TikTok videos using only video text and metadata is similar; 2 models had low accuracy of 50\%, which for a balanced dataset means that it performs equal to a dummy model. 

We surmise that the main reason for this is that audio extraction to text is quite accurate and as most videos include some host(s) speaking about the content, the audio text encapsulates the main idea of the video. 
In contrast, the visual text relies on sampling of visual frames to acquire the  visual text, but this makes several assumptions 1) a video has clear subtitles that are easy to recognize 2) a video displays visual graphics of text that pertain to video's content. 
If a video's visual content had little visual text or did not include subtitles, a classifier had little to base decisions apart from accompanied metadata.
While audio extraction to text suffers from potential limitations such as background music in place of a host's voice, the likelihood is lower that the existence of visual subtitles.
Extracting text from audio also has less likelihood of encountering random audio that may confound the speech-to-text model. 

Therefore, we found that datasets that utilized audio text performed better than datasets that utilized both audio and visual text. 
The only exception was ``TikTok with both visual and audio text" where it actually performed 2\% better than ``TikTok with audio text."
We believe the characteristics of TikTok videos (i.e., increased use of subtitles that complement the audio text of videos over YouTube) may be a factor for why the model could accurately classify ``TikTok with both video and audio text". We expand on the effect of video content characteristics in Section \ref{rq_characteristics}.

\begin{tcolorbox}
\textbf{Findings 1:} Text extraction from videos using audio was highly effective for classifying videos from YouTube and TikTok. Text extraction using video on its own was not effective. However, for TikTok, the combination of text extraction using both video and audio was the most accurate option.
\end{tcolorbox}

GPT-2 and RoBERTa models were most accurate for classifying the videos.
For YouTube videos with audio text, GPT-2 significantly outperformed the other four models by 8-15\%.
RoBERTa was the best performing classifier for two out of the three TikTok datasets.
Roberta had the highest accuracy for ``TikTok with audio text" and ``TikTok with both audio and visual text" with respective accuracies of 93\% and 97\%.
The 97\% RoBERTa achieved for TikTok with both video and audio text was highest accuracy we obtained in all our tests in Table \ref{tab:train_test_models}.



\begin{tcolorbox}
\textbf{Findings 2:} Deep learning models such as GPT-2 and RoBERTa can be utilized to perform classification of video content into relevant and non-relevant user feedback. GPT-2 was the most accurate model for classifying YouTube videos and RoBERTa was the most accurate model for classifying TikTok videos.
\end{tcolorbox}

\begin{table}[]
    \centering
        \caption{Result from Labelling and Classifying the Dataset}
    \begin{tabular}{p{5.6cm}cc} \toprule
    Dataset &  Relevant  &  Irrelevant \\ \midrule
    YouTube Manual Labelling  &  370  & 167 \\ 
    YouTube with audio text classification via GPT-2  & 1691 & 1029  \\
    TikTok Manual Labelling   &  226 & 311 \\
    TikTok with both video and audio text classification via RoBERTa   & 810  &  1672 \\
    Total  &  3097 & 3179 \\ \bottomrule
    \end{tabular}

    \label{tab:relevant_vs_irrelevant_split}
\end{table}

After determining the most accurate models for TikTok and YouTube, we proceeded to classify the rest of the unlabelled dataset using these models.
After classifying the rest, we also took a random sample of 50 videos with their classified labels and performed a round of manual annotation to determine the accuracy of the automated labelling.

We see consistency with the original experiments in Table \ref{tab:train_test_models} in the manual annotation.
``YouTube with audio text" paired with GPT-2 achieved an accuracy of 98\% and ``TikTok with both video and audio text" paired with RoBERTa achieved 100\%. 
We show in Table \ref{tab:relevant_vs_irrelevant_split} the splits for \emph{relevant} and \emph{irrelevant} in the videos. 
In total, we found 3097 videos with relevant information and 3179 videos with non-relevant information for the 20 products in our study.
YouTube videos (61\%) had a higher concentration of requirements elicitation relevant videos compared to TikTok (34\%). 

\begin{tcolorbox}
\textbf{Findings 3:}  Videos from YouTube and TikTok can be used to identify requirements relevant user feedback. Videos from YouTube (i.e., 61\%) are more likely than videos from TikTok (i.e., 34\%) to contain requirements relevant feedback.
\end{tcolorbox}

\subsection{RQ2: What are the main user feedback themes that we can identify?}
We were able to group and cluster the videos based on their user feedback, allowing us to identify the most prominent themes, which are presented in  Table \ref{tab:clustering themes}. 
the themes were generated by reviewing the formed clusters and assigned theme names that represented each cluster accurately.
These themes can serve as a foundation for further refinement and analysis for a company to derive requirement statements.

We found out that every product has videos related to the theme \textbf{Feature Ratings}. 
A \textit{Feature Ratings} provides an overall rating of a product and often highlights specific features that are considered to be strengths or weaknesses. 
For example, \qsd{I hate Firefox but I'm still switching back to it. I can't stand Firefox. Every time I try to use it I just get frustrated by all of the useless features privacy invading telemetry and annoying defaults. If only there was a way to use Firefox without all the junk.}{Firefox}{YouTube} 

Additionally, we observed a consistent theme among software product discussions regarding user feedback on updates and features, which also falls under Feature Ratings. 
Related to the latest updates and features, user often offered suggestions for further improvements. 
For instance, \qsd{Please boost this so that Duolingo can see this I hate this new update so much... You can't jump between topics anymore which is so bad...}{Duolingo}{TikTok}.


A significant number of videos related to \textbf{Bug Reports} provided detailed explanations of the issues they encountered while using the software and may include suggestions for workarounds or solutions. 
To illustrate, \qsd{Fix Discord... Discord App not launching on Windows 10... [To fix, Method 1] Close discord in task manager and restart it. Right click on the taskbar and click on task manager. Right click on the Discord option and click on end task.}{Discord}{YouTube}. When it comes to software products, user-generated videos related to bug reports can be incredibly useful. These videos can provide developers with valuable insight into the issues that users are experiencing with their products.
 
Videos for \textbf{Usage Tutorial} help others with their user experience.
Video content around software products frequently showed hacks and workarounds to help users make the most of their software tools.
This highlights the importance of considering user feedback when designing and updating software products. 
\textit{Usage tutorials} can be extremely useful for companies as they provide insight into how users are interacting with their products and potential areas where usability may be a concern. 
Previous studies have explored converting user feedback into requirements \cite{gebauer2008user, jiang2014user, kengphanphanit2020automatic}, hence an organization can similarly analyze the video content to identify common areas of confusion or difficulty in the usage of their products.

While \textbf{Feature Rating} was a theme found across all the products, it emerged as a particularly prominent theme for phones and computers.
The prevalence of this theme in these product categories could be a result of features playing a crucial role in the purchase decision for these phones and computers. 
This theme was consistently observed across both TikTok and YouTube. 
However, our analysis also revealed that YouTube provides an in-depth review of a product against its competitors, which we labeled as \textit{Matching competition}. 
To illustrate, \qsd{The Pixel gets something iPhone users can only dream of and that is a 48 megapixel telephoto lens with five times optical zoom. Now on iPhone you only have a 12 megapixel telephoto that does three times optical zoom...}{Google Pixel 7}{YouTube}.

Such videos can be incredibly useful for all the related companies, as organizations can learn about the strengths and weaknesses in their product in a highly competitive market. 
With comparisons of similar products, a video can influence prospective customers on purchase decisions. 
A company can unlock highly useful user feedback regarding their own and competitor products if they collect these types of videos. 

A prevalent theme among computers was the evaluation of the performance and design. 
\textbf{Design Ratings} revolve around creators focusing on the aesthetic appeal of these devices, while \textbf{Performance Ratings} often concern the performance quality of these products.
For example, \qsd{There's a few things about the S22 Ultra that just drive me up the freaking war.. this is the most powerful phone Samsung has ever made except it still lags in like the most random places ...}{Samsung Galaxy S22}{TikTok}
The example indicates to other users concerns over the performance of the device while using certain actions, which may be relevant user feedback for Samsung.

Automotive products also exemplified the \textbf{Feature Ratings} theme along with \textbf{Affordability}, \textbf{Performance Ratings}, and \textbf{Modification Suggestions}, which are important aspects of a product that can influence a user's decision-making.
For \textit{Modification Suggestions}, content creators often suggest modifications that other customers and users can add to their product to enhance their experience.
When users and customers express their willingness to enhance their products through modifications, it indicates that they have a clear idea of their desired product features and functionalities.
For example, \qsd{My top 20 Toyota Rav 4 aftermarket upgrades and modifications Mods...I installed [improved] driving lights and I really like the driving lights. The ones that come with it are great, but of course they only work when you have it on low beam and these are of course for the fog lights.}{Toyota Rav 4}{YouTube}
These videos provide opportunities for developers to better understand how customers use their products and identify areas for customization.

Furthermore, \textbf{Repair and maintenance} was also seen as a prominent theme from videos of automotive products.
Video creators frequently discussed the various parts of a car that tend to wear out over time and shared tips on how to avoid and repair them. The information provided in these videos can be a valuable resource for automotive companies looking to improve the quality of their products.
By analyzing the patterns of wear and tear identified by these video creators, companies can gain insights into how their products are performing over time. They can use this information to make improvements to the design and construction of their products to enhance their longevity and durability.



\begin{tcolorbox}
\textbf{Findings 4:} User feedback themes can be generated from videos from YouTube and TikTok. 
These user feedback themes not only represent important aspects about products for companies to consider, but also represent relevant user feedback that companies can further refine into requirements in a subsequent step.
\end{tcolorbox}

\begin{table}[]
\caption{Requirement Relevant Themes}
\begin{tabular}{p{2cm}p{4.3cm}p{1.5cm}}
\toprule
Theme                     & Description                                      & Number of Products (out of 20) \\ \midrule
Feature Ratings           & Praise/criticism of product features             & 20                       \\
Matching Competition      & Comparison with other competitor products        & 13                       \\
Performance Ratings       & Praise/criticism of performance of the products  & 8                        \\
Modifications Suggestions & Suggestions for tools/upgrade                    & 5                        \\
Bug Report                & Bugs and issues of products                      & 4                        \\
Repair \& Maintenance     & Videos related to fixing and preserving          & 3                        \\ 
Design Ratings            & Design evaluation                              & 3                        \\
Affordability             & Cost prospects of the products                   & 3                        \\
Usage Tutorials           & Tutorials for other user to help use the product & 2                        \\
\bottomrule
\end{tabular}

\label{tab:clustering themes}
\end{table}

\begin{table*}[]
\centering
\caption{Video Content Statistics}
\begin{tabular}{lp{1cm}p{0.7cm}p{1cm}p{1cm}p{1.3cm}p{1.3cm}p{1.1cm}p{1.1cm}p{1.3cm}p{1.1cm}p{1.1cm}p{1.1cm}}
\toprule
Platform & Product & Avg. Sec. & Views Per Video & Audio Words Per Video & Uniq. Audio Words Per Video & Audio Words Per Sec. & Uniq. Audio Words Per Sec. & Visual Words Per Video & Uniq. Visual Words Per Video & Visual Words Per Sec. & Uniq. Visual Words Per Sec. \\ \midrule
\multirow{5}{*}{YouTube} &  Software & 560 & 0.20M & 808 & 253 & 1.4 & 0.5 & 546 & 231 & 1.0 & 0.4\\
& Phone & 555 & 1.48M & 934 & 334 & 1.7 & 0.6 & 232 & 111 & 0.4 & 0.2\\
& Laptop & 480 & 0.20M & 1,356 & 465 & 2.8 & 1.0 & 319 & 165 & 0.7 & 0.3\\
& Car & 450 & 0.22M & 946 & 317 & 2.1 & 0.7 & 176 & 91 & 0.4 & 0.2\\
& Total & \textbf{509} & \textbf{0.50M} & \textbf{986} & \textbf{333} & \textbf{1.8} & \textbf{0.7} & \textbf{313} & \textbf{146} & \textbf{0.6} & \textbf{0.3} \\ \midrule
\multirow{5}{*}{TikTok} 
& Software  & 32 & 1.11M & 79 & 49 & 2.5 & 1.5 & 216 & 90 & 6.7 & 2.8 \\
& Phone  & 36 & 1.84M & 76 & 49 & 2.0 & 1.4 & 98 & 42 & 2.7 & 1.2\\
&  Laptop & 39 & 0.21M & 75 & 48 & 1.9 & 1.2 & 119 & 46 & 3.0 & 1.2 \\
& Car  & 37 & 1.08M & 83 & 53 & 2.2 & 1.4 & 61 & 28 & 1.6 & 0.8 \\

& Total  & \textbf{36} & \textbf{1.07M} & \textbf{79} & \textbf{50} & \textbf{2.4} & \textbf{1.5} & \textbf{120} & \textbf{50} & \textbf{3.3} & \textbf{1.4} \\

\bottomrule
\end{tabular}

\label{video_stats}
\vspace{-10pt}
\end{table*}

\subsection{RQ3: How do the different social media platforms and their video content affect the user feedback?} \label{rq_characteristics}
We observed in RQ1 how the number of videos containing relevant user feedback that can be later refined into requirements differed significantly between YouTube and TikTok.
Table \ref{video_stats} illustrates some of the differences between the audio and visual texts between YouTube and TikTok videos.
In particular, we found that YouTube videos on average were at least 15 times longer than TikTok videos. 
It is then not surprising that higher percentage of YouTube videos would be afforded time to cover content related to feature ratings, bug reports, and discuss missing features that competitor products provide. 
It is more difficult to cover these topics in a short span of 33 seconds, which is the average video span for TikTok.

However, we also noticed that on average TikTok video covered more words and unique words per second than YouTube. 
In particular, a TikTok video typically covered more than 2 times more unique words per second than YouTube. 
We believe this phenomenon was occurring largely because content creators have to squeeze more content in a shorter time frame. Hence, the audio is likely sped up to attract user attention. 

Despite having shorter videos on TikTok, TikTok videos attract on average 2 times more views than the YouTube videos. This perhaps is not too surprising as TikTok videos are much shorter, so the theoretical videos viewed per hour watched for a typical user is likely also much higher than on YouTube. 
However, the higher view count is still a non-negligible characteristic that organizations should consider as higher view count may result in a bug report or feature rating becoming more influential.

TikTok videos on average had 5 times more visual texts per second than on YouTube and the number of unique visual texts per second was also significantly more. 
The high number of visual text on TikTok likely stems from the increased use of video subtitles on TikTok than YouTube so the visual texts often complement the audio of a video.
The complementary factor between visual and audio text is potentially another reason why for TikTok the dataset with both visual and audio text had the highest accuracy among all TikTok datasets.
We contrast this with YouTube videos where visual subtitles are less likely and the text that appear may actually confound the classifying models.

These differences may have contributed to the different results from our classification models. 
We used GPT-2 medium and RoBERTa base for our study, but GPT-2 medium is a significantly larger model with greater number of model parameters than RoBERTa base (i.e., Roberta = 125M vs GPT2=345M).
The condensed nature of TikTok videos, generates less noise in the data requiring fewer model parameters.
Smaller models tend to fit better to smaller data as they are less prone to overfitting, due to the fewer parameters.
Our results largely reflect this expectation as RoBERTa was the most accurate model for 2 out of 3 TikTok datasets and GPT-2 was the most accurate for 3 out of 6 datasets.

\begin{tcolorbox}
\textbf{Findings 5:} 
YouTube has more videos with relevant user feedback than TikTok likely due to the factors of longer video and more in-depth discussions about each product. 
TikTok videos on average are more than \emph{15 times} shorter than YouTube with much more condensed content, but has on average \emph{double} the number of average views. 
Due to the frequent use of subtitles in TikTok videos, visual text can assist audio text in determining relevant user feedback that can be used later for developing requirements. 
These characteristics also affect the accuracy of various deep learning models that work for each platform. 
\end{tcolorbox}


\section{Discussion}
Our work indicates the potential to improve the practices of CrowdRE by utilizing valuable information present in video content.

\subsection{Videos: A Source of Requirements Relevant User Feedback}
TikTok and YouTube offer an interactive and immersive space for users to engage with the ``crowd" through content creation. 
These platforms allow viewers to interact with the creator through likes, comments, shares, and reactions.
In this work, we explore how videos extracted from these two social media can be used to identify requirements relevant user feedback.

For example, \qsd{The new Duolingo update is seriously messing me up I can't even get back into the lessons I was actively working on. Please revert it... Goodbye Duo it was fun. ... Also note that this person has super Duolingo which means they pay for a subscription}{Duolingo}{TikTok}
This exemplifies how a paying user is leaving the product due to a recently introduced update on Duolingo, which has resulted in a series of bugs on the platform.
For Duolingo, if they considered such video, the actionable requirement would entail reverting the update or fixing the bug to not impede the user's experience. 
An organization that seeks to reduce user attrition may benefit from these bug report insights, and utilize them to develop requirements that developers could implement.

Furthermore, people often discuss about the problems they encounter while using a certain product, \qsd{Great phone some bugs Google Pixel 7 Pro... I've just had quite a few instances where things will just randomly freeze up like apps will just get stuck or I'll get stuck on just a black screen and can't get out of it... things won't always work all the time which is kind of frustrating...}{Google Pixel 7}{TikTok}
Although bug report are a common issue for products in textual feedback (e.g. forums, app reviews), videos from YouTube offer extremely rich details about issues \cite{khanresearching}.

The videos from YouTube and TikTok offer a level of detail into problems and issues that companies can reference to understand underlying problems. 
For example, an app review may just include textual description about an issue \cite{maalej2015bug}, a Reddit post may include textual description along with a screenshot, but a video may include a short clip about how a bug was triggered or the outcome of the bug that the company can re-watch when they are creating actionable requirements.
Since these videos are quite popular on TikTok and YouTube, they may influence potential new and current users with the perceived honest and objective opinions.
Therefore, our work highlights the greater importance that organizations should place on analyzing the video based content for requirements relevant user feedback. 

\textbf{\textit{Example use case:}} We present an example scenario regarding how organizations can leverage the approach proposed in this study.
Duolingo is one of the premier language learning applications in the world.
If they manage their online user feedback, they could search and download for Duolingo related videos from video based social media like YouTube and TikTok.
Upon converting the audio text and visual text from each video, the organization would be left with a series of videos with their corresponding audio and visual text.
The organization could then run a RoBERTa model for all its TikTok video to identify the requirements relevant user feedback.
Subsequently, the organization could use a topic modeling model, like BERTopic, to identify the various types of feedback.
If the organization wants to identify bugs, they could expect to find bug related topics from the topic modeling.
For instance, there could be multiple videos covering the issues related to the Duolingo update. 
If the problematic update was covered by multiple videos, the organization would find a common theme that describes the issue.
The next step for the organization would be eliciting actionable requirement(s), maybe in the form of user stories, to resolve the user concern.
Eliciting the actionable requirement is outside the scope of this work, but an organization can extend our approach in a further step.
The work left for a product person to create a user story for their issue tracker is quite straight forward, as they could already infer the type of issue (i.e., bug, feature request, etc) and the content from a video is generally quite explicit about the specific issue.

\subsection{Implications for Practitioners}
Our findings suggest a number of implications for practitioners who can incorporate the user feedback from the video contents as part of their requirement generation process.
An organization may learn about how users are rating their products in terms of features, design, specifications, and performance through the video contents itself. 
A company looking to improve their products in comparison to other competitors can take advantage of the competitor analysis found in these videos. 
The software products like Chrome, Firefox, Notion, Discord videos often contain feature and user experience related discussion, that may be beneficial for organizations to consider before rolling out a new feature. 
Products related to automotive companies have the potential to learn about user concerns related to repair, maintenance, and efficiency etc. 
Many of the videos further express the consumers feedback about affordability, customization and modification suggestions. 
Organizations have been analyzing YouTube videos for improving marketing and advertising purposes for a while now \cite{kousha2012role}.

The cost for an organization to adopt our approach is for the most part minimal.
They would need to build a web scraping pipeline to download the videos from YouTube and TikTok, but once its built, they can repeatedly use it. 
The main cost would likely come from extraction of visual text as OCR extraction using third-party subscriptions may be expensive, but there are alternative open-source tools that are available.
Processing a software organization's sample batch of 100 TikTok videos would take approximately 15 minutes for audio text extraction and 1.5 hours for visual text extraction on 2 x Intel E5-2683 v4 Broadwell @ 2.1GHz and a P100 16G RAM.
Once an organization implements data analysis of user feedback similar to our approach they can collect the main user feedback themes.
The final step would involve having an employee such as product manager or technical lead to parse the user feedback themes into actionable requirements, but this should be straightforward task.
For example, in the sample content regarding Duolingo's flawed new update, a product manager can quickly see that at minimum the organization should roll back changes to a previous version. 
Otherwise, the organization should implement bug fix to allow users to access current lessons.
Hence, the actual real dollars cost to an organization to use our approach is limited, and the organization could realize benefits of obtaining user feedback themes based on insights from the crowd. 

Users are engaging in video content creation on a regular basis, providing various feedback about the products.
Our study has potential to influence  industry requirements and product management practices, as practitioners can gain valuable insights about user behavior and concerns.

\subsection{Implications for Research}
We believe our research has several important implications for researchers and future studies. 
First, we identified the efficacy of leveraging videos from two large social media platforms for identifying product requirements relevant user feedback. 
Social media, particularly video based on social media platforms, has exploded in popularity in recent years and their growth across different demographics provides new and important new sources for CrowdRE. 
Other large platforms are joining this foray, with Meta's Instagram Posts and Reels and Twitter's Vertical being the significant alternatives. Future research should explore these alternative patterns and study whether these patterns have characteristics that inhibit or enable relevant information for products.

Second, another area of future work is in the methodological domain. 
In our approach, we tried to focus on larger, more pronounced visual texts in videos as opposed to every possible text that may appear in a single frame, but future work should consider other approaches to analyzing the visual text. 
Other approaches may help to increase the usefulness of the visual text for identifying user feedback, especially on TikTok where we noticed that there were more visual texts in general. 
There is also the potential for researchers to explore identifying other information from the visual aspect of videos.
Potential areas may involve automated interpretation of the visual content in a video and converting that into text.

While we tried to focus on user generated content, by filtering out official product account videos, it does not fully clear the dataset of potential sponsored or promotional types of videos. 
Future research could further separate these promotional content, perhaps through a classification filtering stage similar to the models used in our study, and explore the user generated content in more detail.
Additional work from research could include automatic detection about actual feature requests or bugs as this could assist practitioners in identifying requirements from the videos.

Finally, future work can involve correlating video content with other accompanying characteristics in a video such as the number of likes, the number of comments, as well as the content of those comments. 
Previous work has already explored the usefulness of video comments \cite{karras2021potential}, but utilizing both the video itself and its accompanying data could prove even more effective for interpreting user feedback that may be refined into requirements through subsequent steps.

\section{Threats to Validity}

\subsection{Construct Validity} Construct validity relates to whether we measured what we intended to measure. In our case, one threat relates to our manual labelling of ``relevant" versus ``irrelevant". There could be subjective bias introduced in this labelling, but we tried to mitigate this through a definition of this concept from literature and two coders who have in-depth experience in requirements concepts and pair coding. We used our Cohen's kappa and agreement levels as a measure for our reliability. 

\subsection{External Validity} In terms of external validity, there is the limitation that our study may not generalize to other video platforms or other software products. However, we tried to mitigate this issue by studying two leading video based social media platforms and exploring 20 leading products from 4 major industries. Therefore, we anticipate that videos from other software products on TikTok or YouTube will produce similar results.

\subsection{Internal Validity} 
Our conclusions about the visual text extraction could be limited by our project budget in terms of optical character recognition (OCR) extraction.  The ``Google Vision" API had superior performance over the HuggingFace API, but the cost of the ``Google Vision" API at \$1.50 per 1000 frames was a constraining factor. Also, videos are high-redundancy media, which increases processing time, even with the algorithmic frame sampling we employed.  Increasing the sampling rate will decrease the computation run-time, at the expense of the information-loss rate. We chose parameters to sample at an minimum rate of 1.5s/frame and 2.5s/frame for TikTok and YouTube, respectively.  This may have caused missed frames containing pertinent information.

\section{Conclusion}
Video based social media platforms generate a wide range of discussions regarding different products, and analyzing these platforms have become a popular CrowdRE practice.
In this study, we explored how can we leverage videos from two of the most popular video based social media platforms: TikTok and YouTube, for identifying requirements relevant user feedback that may be refined later into requirements. 
We examined 20 different products across a range of industries and used NLP and machine learning techniques to analyze the audio and visual content from the two platforms.
We found that deep learning models such as GPT-2 and RoBERTa are effective in classifying video content into relevant and non-relevant user feedback, and clustering techniques can be used to identify user feedback themes.
Popular themes that emerged include user feedback about feature ratings, bug reports, performance and efficiency issues about the software and other consumer products. 
As videos continue to gain popularity as a medium of communication and content creation, organizations can benefit from leveraging this data source to gain insights into user needs for their products.


\bibliographystyle{IEEEtran}
\bibliography{main}

\end{document}